\documentclass[prl,aps,twocolumn,superscriptaddress,showpacs]{revtex4}
\usepackage{amsmath,amssymb,graphicx}
\usepackage{pxfonts}
\usepackage{graphicx}
\usepackage{color}
\usepackage{float}
\begin{document}

\title{Spatially and time-resolved imaging of transport of indirect excitons in high magnetic fields}

\author{C. J. Dorow}
\author{M. W. Hasling}
\author{E. V. Calman}
\author{L. V. Butov}
\affiliation{Department of Physics, University of California at San Diego, La Jolla, California 92093-0319, USA}

\author{J. Wilkes}
\affiliation{School of Physics and Astronomy, Cardiff University, Cardiff CF24 3AA, Wales, United Kingdom}

\author{K. L. Campman}
\author{A. C. Gossard}
\affiliation{Materials Department, University of California at Santa Barbara, Santa Barbara, California 93106-5050, USA}

\begin{abstract}
We present the direct measurements of magnetoexciton transport. Excitons give the opportunity to realize the high magnetic field regime for composite bosons with magnetic fields of a few Tesla. Long lifetimes of indirect excitons allow the study kinetics of magnetoexciton transport with time-resolved optical imaging of exciton photoluminescence. We performed spatially, spectrally, and time-resolved optical imaging of transport of indirect excitons in high magnetic fields. We observed that increasing magnetic field slows down magnetoexciton transport. The time-resolved measurements of the magnetoexciton transport distance allowed for an experimental estimation of the magnetoexciton diffusion coefficient. An enhancement of the exciton photoluminescence energy at the laser excitation spot was found to anti-correlate with the exciton transport distance. A theoretical model of indirect magnetoexciton transport is presented and is in agreement with the experimental data. 
\end{abstract}


\date{\today}

\maketitle

\section{I. Introduction}

An indirect exciton (IX) is a bosonic (quasi)particle consisting of a bound electron (e) and hole (h) confined to spatially separated quantum well (QW) layers [Fig.~1(a)]. The spatial separation of the electron and hole reduces their wave-function overlap, leading to IX lifetimes several orders of magnitude longer than those of regular direct excitons. IXs also form oriented dipoles and have repulsive interactions that facilitate disorder screening. The long IX lifetimes and ability to screen disorder enable long-range transport of IXs before recombination, with transport distances reaching up to hundreds of micrometers~\cite{Hagn95, Butov98, Negoita99, Larionov00, Butov02, Voros05, Gartner06, Ivanov06, Hammack09, Lasic10, Alloing11, Alloing12, Lasic14}. 

Several transport phenomena have been observed in systems of IXs, including the inner ring in the IX emission pattern~\cite{Butov02, Ivanov06, Hammack09, Alloing12, Stern08, Ivanov10, Kuznetsova12}, laser-induced IX trapping~\cite{Hammack06, Hammack07, Gorbunov11, Alloing13}, IX localization-delocalization transitions in random~\cite{Butov02, Ivanov06, Hammack09}, periodic \cite{Remeika09, Remeika12}, and moving~\cite{Winbow11, Hasling15} potentials, IX spin transport \cite{Leonard09, High13}, and coherent IX transport with suppressed scattering~\cite{High12, Alloing14}. Long-range IX transport has also enabled the realization of several excitonic devices including excitonic ramps~\cite{Hagn95, Gartner06, Leonard12, Dorow16}, conveyers~\cite{Lasic10, Lasic14, Winbow11, Hasling15}, and transistors~\cite{High07, High08, Grosso09, Kuznetsova10, Andreakou14}. 

\begin{figure}[htbp]
\centering
\includegraphics[width=\linewidth]{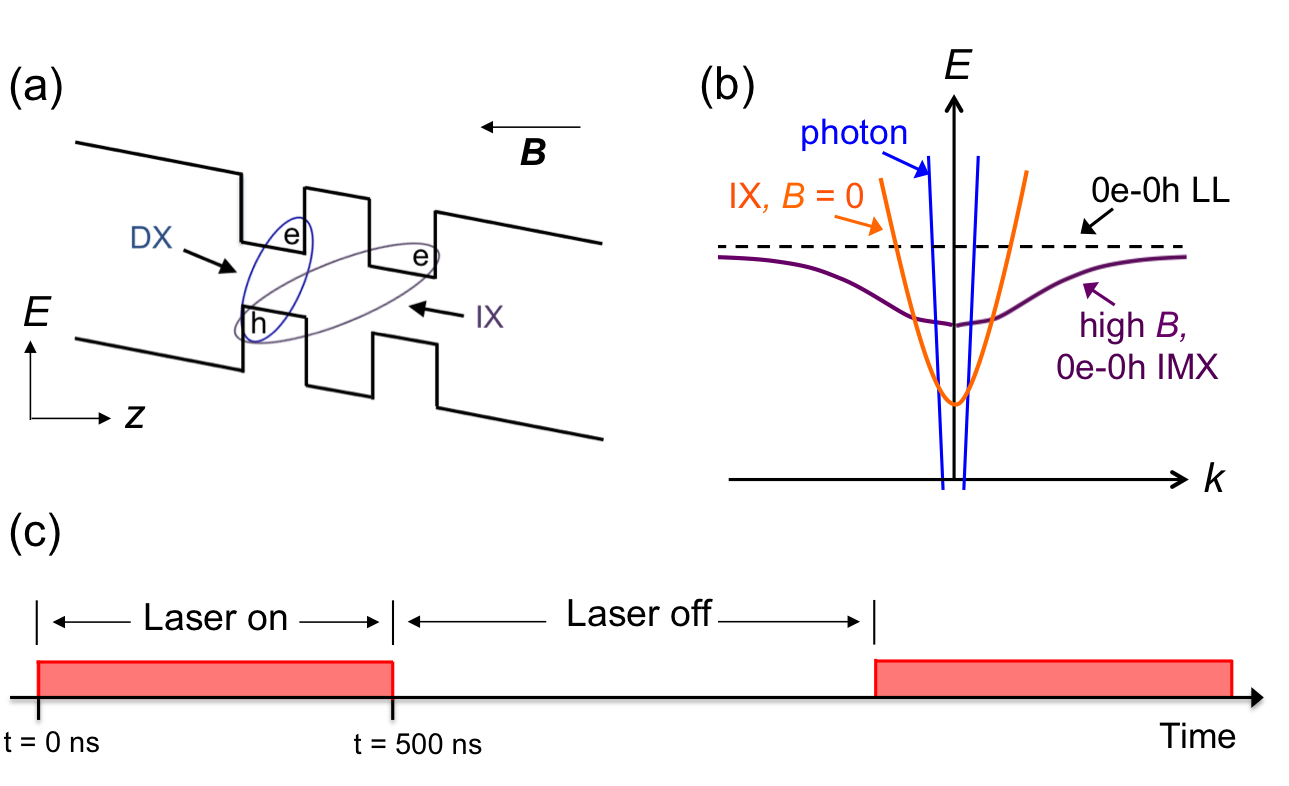}
\caption{(a) CQW band diagram. (b) Dispersion for indirect exciton (IX) in zero magnetic field (orange). Dispersion for indirect magnetoexciton (IMX) formed from electron (e) and hole (h) in zeroth Landau levels (LLs) in high $B$ (purple). Sum of the e and h LL energies (black dashed line). Photon dispersion (blue). (c) Schematic diagram of the rectangular laser excitation pulse profile. Time $t = 0$ ns corresponds to the onset of the laser pulse. The pulse width $\tau_{\rm width} = 500$ ns and pulse period $\tau_{\rm pulse} = 1.3$~$\mu$s.}
\end{figure}

A system of IXs also provides a unique opportunity to study transport of cold bosons in the high magnetic field regime. While transport of fermions in high magnetic fields has shown remarkable properties \cite{QHE}, studies of bosons in the high magnetic field regime have remained elusive to experimenters due to the magnitude of the magnetic fields required. The high magnetic field regime is achieved when the cyclotron energy is comparable to the binding energy of the boson constituents, which requires $B \sim 10^6$~Tesla for atoms. In contrast, the high magnetic field regime for excitons is achievable in a laboratory, requiring only a few Tesla, due to their small masses and binding energies~\cite{Lozovik02}. In addition to enabling the high magnetic field regime, IXs possess several other advantageous properties for studying transport of cold bosons in high magnetic fields: Long IX lifetimes enable transport distances that are large enough to be observed with optical imaging; IXs can cool to low temperatures, below the temperature of quantum degeneracy, due to their long lifetimes and fast energy relaxation in QWs~\cite{Butov01}; The density of photoexcited e--h systems can be controlled by the laser excitation, which allows the realization of virtually any Landau level (LL) filling factor. 

\begin{figure*}[htbp]
\centering
\includegraphics[width=\textwidth]{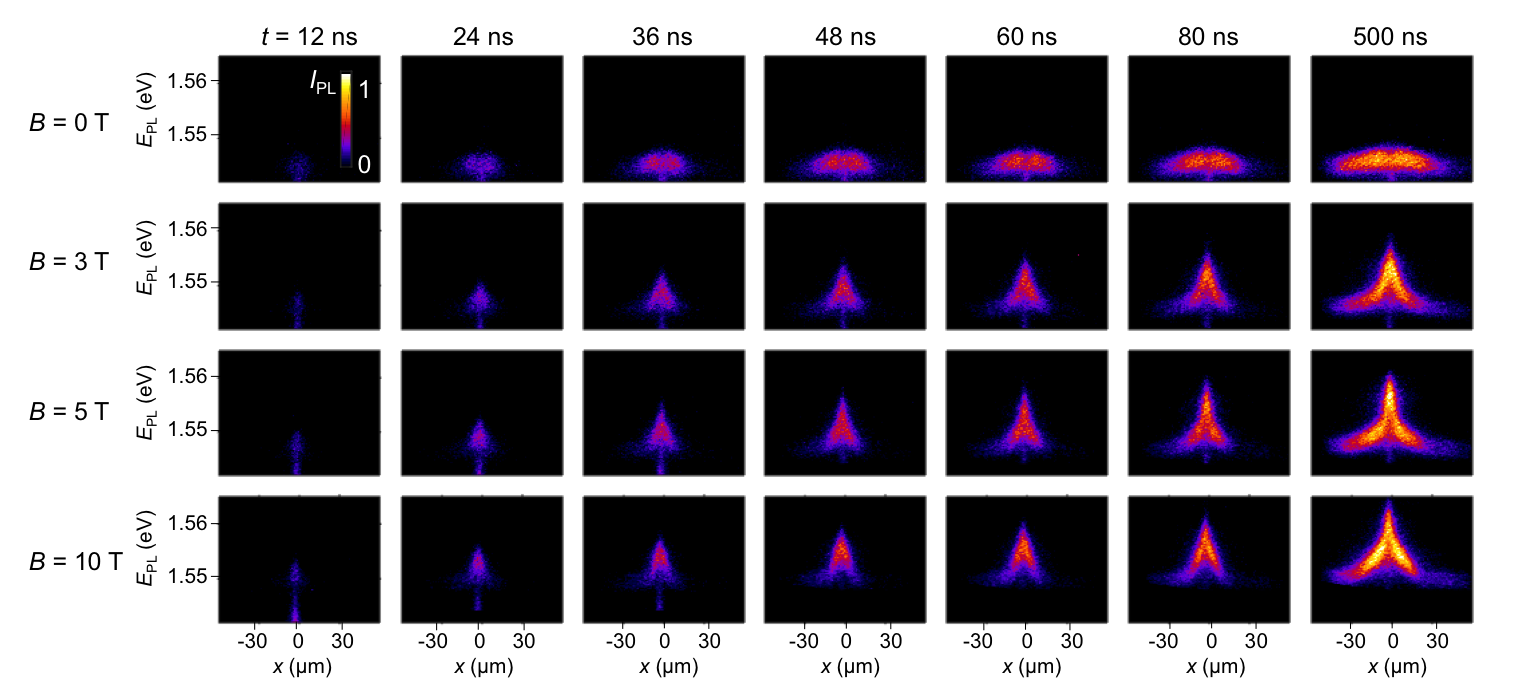}
\caption{$x$-energy IMX photoluminescence (PL) images taken during the laser excitation pulse. Images are taken for several delay times $t$ after the onset of the laser pulse and several magnetic fields. Each image is integrated over a time window of $\delta t = 4$~ns. $T_{\rm bath} = 1.5$~K and average $P_{\rm ex} = 670$~$\mu$W for all data. Laser excitation is centered around $x$ = 0.}
\end{figure*}

An exciton in the high magnetic field regime, a magnetoexciton (MX)~\cite{Gor'kov68, Lerner80, Kallin84, Lozovik02}, is composed of an electron and a hole in LL states. The MX dispersion is defined by a coupling between the MX center-of-mass motion and internal structure: an electron and a hole of a MX are forced to travel with the same velocity and produce on each other a Coulomb force that is balanced by the Lorentz force. This causes MXs with momentum $k$ to acquire an in-plane electric dipole $r_{\rm eh} = kl_{\rm B}^2$ in the direction perpendicular to $k$, where $l_{\rm B} = \sqrt{\hbar c / (eB)}$ is the magnetic length. The coupling between $r_{\rm eh}$ and $k$ allows the MX dispersion $E(k)$ to be calculated from the Coulomb potential between the electron and the hole as a function of $r_{\rm eh}$. At small $k$, the MX dispersion is parabolic and can be described by an effective MX mass that grows with $B$. Figure~1(b) shows the transformation of the MX dispersion from $B = 0$ to high $B$ for a $0_{\rm e} - 0_{\rm h}$ MX state (MX composed of an electron and hole each in the zeroth LL). Several collective states and transport phenomena have been predicted for gases of MXs including a paired Laughlin liquid \cite{Yoshioka90}, an excitonic charge-density-wave state \cite{Chen91}, a condensate of MXs \cite{Kuramoto78, Lerner81}, the exciton Hall effect~\cite{Dzyubenko84, Paquet85}, superfluidity~\cite{Kuramoto78, Lerner81, Paquet85, Imamoglu96}, and localization~\cite{Dzyubenko95a}.

The MX properties can be probed by photoluminescence (PL) techniques. In GaAs QW structures, free 2D MXs can recombine radiatively when they are composed from electrons and holes at LLs with $N_{\rm e} = N_{\rm h}$, where $N$ is the LL number of the electron and hole, their spin projections on the $z$ direction is $J_z = \pm 1$, and their momentum $k$ lies within the intersection between the MX dispersion surface $E_{\rm MX}(k)$ and the photon cone $E = \hbar k c / \sqrt{\varepsilon}$ [blue curve in Fig.~1(b)], called the radiative zone~\cite{Feldman87, Hanamura88, Andreani91, Maialle93} ($\varepsilon$ is the dielectric constant). Free MXs with $N_{\rm e} \ne N_{\rm h}$, $J_z = \pm 2$, or $k$ outside the radiative zone remain dark.

In earlier studies of excitons in high magnetic fields, excitons and deexcitons in dense e-h plasmas in single QWs were observed~\cite{Butov91, Butov92}. However, short lifetimes of excitons in single QWs did not allow the achievement of low exciton temperatures or long-range exciton transport before recombination. Recently, the transport properties of cold MXs were addressed with indirect magnetoexcitons (IMXs) in coupled QWs (CQWs) by imaging of IMX emission cloud in cw experiments~\cite{Kuznetsova17}. In these experiments, long-range transport was evidenced for IMXs composed of electrons and holes in the lowest LLs, however no time-resolved measurement of the IMX cloud expansion directly showing IMX transport was performed. The magentic field effect on IX energies~\cite{Butov95, Dzyubenko96, Butov99, Butov01a, Kowalik-Seidl11, Schinner13}, dispersion relations~\cite{Lozovik97, Butov01b, Lozovik02, Wilkes16, Wilkes17}, and spin states~\cite{Gorbunov13, High13} have also been studied. In this work, we present the first direct measurements of IX transport in high magnetic fields, achieved with spatially, spectrally, and time-resolved optical imaging. 

\begin{figure}[htbp]
\centering
\includegraphics[width=\linewidth]{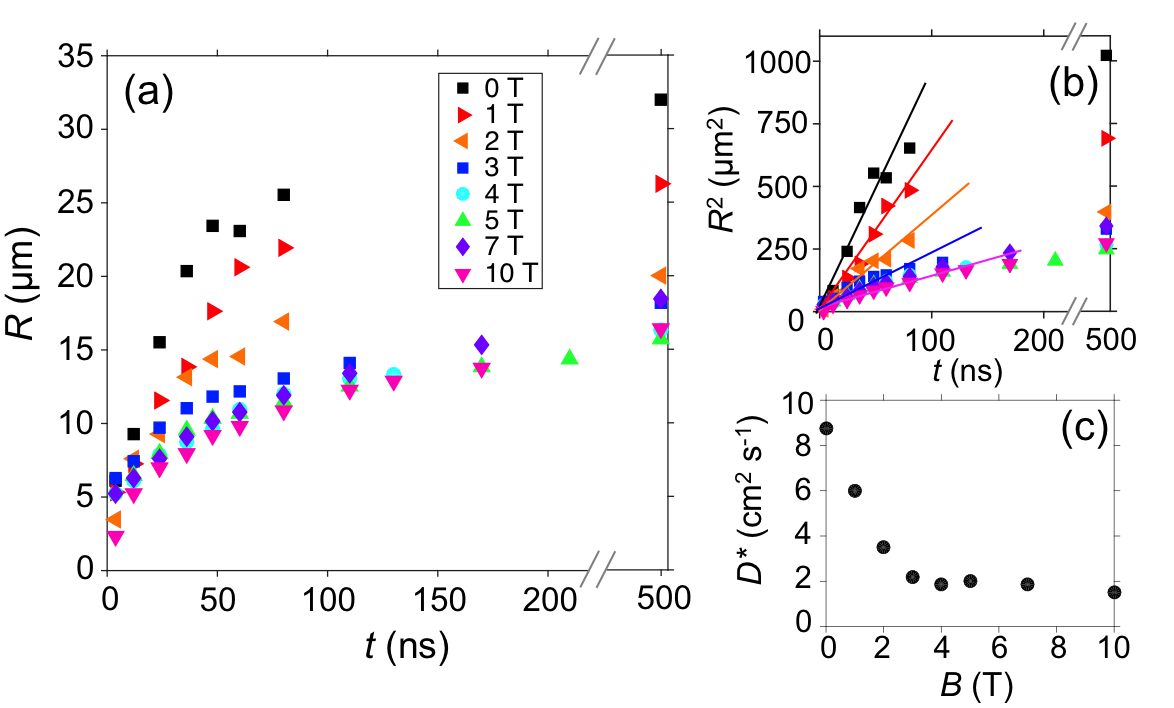}
\caption{(a) IMX transport radius $R$ vs. delay time $t$ for $B = 0$ to 10 T. (b) $R^{2}$ vs. $t$ for $B = 0$ to 10 T. (c) Measured IMX diffusion coefficient vs. $B$ (see text). $T_{\rm bath} = 1.5$~K and average excitation power $P_{\rm ex} = 670$~$\mu$W for all data.}
\end{figure}

\section{II. Experiment}

IXs were studied in a $n^+ - i - n^+$ GaAs CQW grown by molecular beam epitaxy [Fig.~1(a)]. The $i$-region is a pair of 8-nm GaAs QWs separated by a 4-nm Al$_{0.33}$Ga$_{0.67}$As barrier, all surrounded by 200-nm Al$_{0.33}$Ga$_{0.67}$As layers. The spacing between the electron and hole layers derived from the IX energy shift with applied voltage is $d \sim$ 11.5 nm, close to the nominal distance between the QW centers. The $n^+$ layers are Si-doped GaAs with $n_{Si}$ = $5 \times 10^{17}$~cm$^{-3}$. The indirect regime, where the IX is the lowest energy state, was achieved by applying voltage $V = 1.4$~V across the $n^+$ layers. 

Time-resolved optical imaging using a pulsed laser excitation was performed. IXs were photo-generated with a 658 nm laser with a pulse duration of $\tau_{\rm width} = 500$~ns and pulse period of $\tau_{\rm pulse} = 1.3$~$\mu$s with an edge sharpness of $\sim 1$~ns [Fig.~1(c)]. The period and duty cycle were optimized to allow the IMX PL image to reach equilibrium during the laser excitation and decay between laser pulses. The laser was focused to a $R_0 = 4.5$~$\mu$m half-width-half-maximum (HWHM) spot. Images were integrated over 4~ns windows ($\delta t = 4$~ns) and taken for several delay times $t$ after the onset of the laser pulse, defined such that a delay time $t$ corresponds to an image taken during time $t - \delta t$ to $t$.

The images were captured with use of a liquid-nitrogen-cooled CCD coupled to a PicoStar HR TauTec time-gated intensifier. The PL was passed through a spectrometer with a resolution of 0.18~meV before entering the intensifier and CCD in order to obtain spectral resolution. The spectrally- and time-resolved imaging enables the direct measurement of the evolution of the IX PL intensity and energy as a function of delay time $t$. All measurements were performed at $T_{\rm bath} = 1.5$~K with applied magnetic fields up to 10~T oriented perpendicular to the CQW plane.

\begin{figure}[htbp]
\centering
\includegraphics[width=\linewidth]{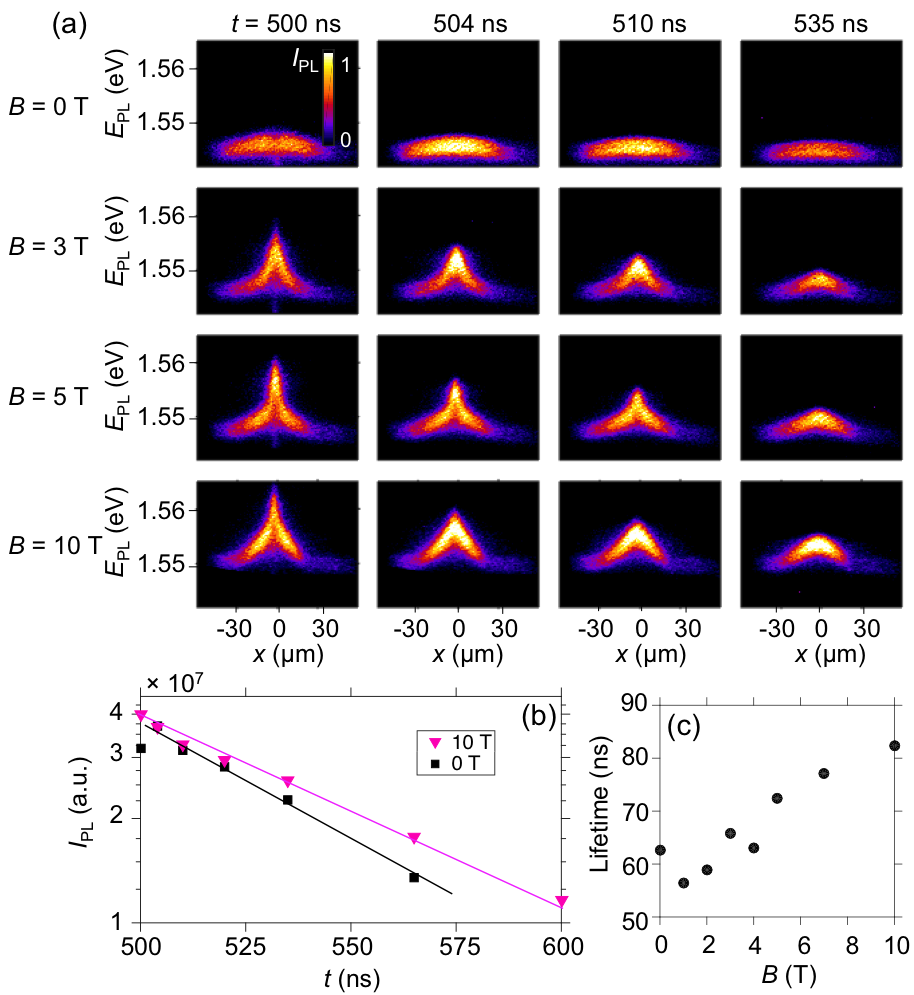}
\caption{IMX decay after laser pulse termination. (a) $x$-energy IMX PL images taken during laser pulse ($t = 500$~ns) and after laser pulse termination ($t > 500$~ns) for several magnetic fields. (b) Total IMX PL intensity (integrated over all $x$ and $E$) vs. $t$ for $B = 0$ and 10~T. (c) Measured IMX lifetime vs. $B$. $T_{\rm bath} = 1.5$~K and average $P_{\rm ex} = 670$~$\mu$W for all data.}
\end{figure}

Figure~2 shows a sample of $x-$energy images of the IMX PL spanning over a range of delay times $t$ and several magnetic fields. All images in Fig.~2 are taken during the laser excitation ($t < \tau_{\rm width}$) with the laser centered around $x = 0$. The magnetic field has an effect on both the IMX transport distance and energy; the IMX PL signal morphs to a chevron shape with increasing magnetic field. The expansion of the IMX cloud due to IMX transport is directly observed over time and is characterized by a transport radius, $R$, where $R$ is the HWHM of the spectrally integrated IMX emission. Figure~3(a) shows $R$ vs. $t$ for each $B$. With increasing magnetic field, the IMX transport slows down, as seen by the flattening of the slope of the curves $R(t)$ with increasing $B$. 

\begin{figure*}[htbp]
\centering
\includegraphics[width=\textwidth]{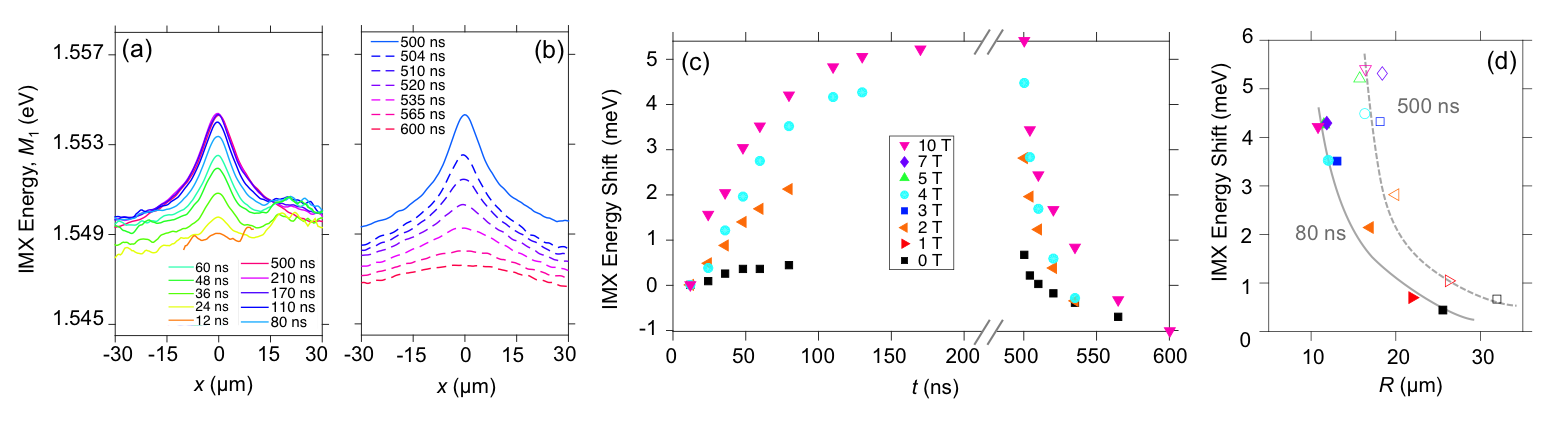}
\caption{IMX energy presented by first moment of IMX PL $M_{\rm 1}(x)$ for different delay times during (a) and after (b) laser pulse. $B = 5$~T. (c) IMX energy shift $M_{\rm 1}(t) -  M_{\rm 1}(t = 12$~ns) at $x = 0$. (d) IMX transport radius $R$ vs. IMX energy shift at $x = 0$ for several $B$ at two delays: $t = 80$~ns (filled markers) and $t = 500$~ns (open markers). $T_{\rm bath}$ = 1.5 K and average $P_{\rm ex} = 670$~$\mu$W for all data.}
\end{figure*}

The slowing down of the IMX transport with increasing $B$ is consistent with the IMX effective mass, $M(B)$, enhancement. The latter was observed in earlier experiments~\cite{Butov01b} and derived in earlier theoretical works~\cite{Lozovik02, Wilkes16, Wilkes17}. For instance, it was shown that for the IXs in the GaAs CQW, the application of magnetic field $B = 10$~T results in the enhancement of $M(B)$ by more than 5 times~\cite{Butov01b, Lozovik02, Wilkes16, Wilkes17}. The IMX mass increase affects both the IMX diffusion and drift since both the exciton diffusion coefficient $D$ and exciton mobility $\mu_{\rm x}$ are inversely proportional to the mass~\cite{Ivanov02}.

The time-resolved imaging of IMX cloud expansion enables estimation of IMX transport characteristics. Figure~3(b) shows that at the initial delay times, $R^{2}$ grows nearly linearly with $t$. Fitting to the slope by $R^{2} \sim R_{0}^{2} + D^{*} t$ gives an estimate of the "effective" IX diffusion coefficient $D^{*}$, which encapsulates both $D$ and $\mu_{x}$. A quantitative description of $D^{*}$ is presented in Section III. The values of $D^{*}$ estimated from this experiment are plotted in Fig.~3(c) and are found to decrease with increasing $B$, quantifying the IMX transport suppression in high $B$. 

We note also that at large delays the slope of $R^2(t)$ reduces. This is consistent with the theory (Section~III) and is mainly due to the finite IX lifetime, $\tau$, which limits the exciton transport distance by $R \sim \sqrt{D \tau}$.

The kinetics of the IMX decay after the laser pulse termination is presented in Fig.~4. The decay of the IMX PL intensity after the laser pulse termination can be fit to extract the IMX lifetime, as seen if Fig.~4(b), where $I_{\rm PL}$ is the total PL intensity integrated over all energies and $x$. The IMX lifetime is observed to increase with $B$ [Fig.~4(c)], in agreement with the earlier spatially integrated measurements~\cite{Butov99}. The IMX dispersion flattening with increasing $B$ [Fig.~1(b)] reduces the energy width of the radiative zone, thus contributing to the IMX lifetime increase.

The evolution of the IMX PL energy is presented in Fig.~5. Figure~5(a) and (b) show the IMX average PL energy $M_{\rm 1}(x) = \int E I(x,E) dE / \int I(x,E) dE$ as a function of delay time $t$ for $B = 5$~T. During the laser excitation pulse, the IMX PL blue shifts over time [Fig.~5(a)]. This can be explained as follows. IMXs are oriented dipoles, with a built-in dipole moment $ed$, and therefore interact repulsively. Thus, IMX energy increases with density. IMXs are photo-excited and their density grows with time after the onset of laser excitation, leading to a blue shift of the IMX emission line with time. Similarly, IMX emission line red shifts after the laser excitation is terminated [dashed lines in Fig.~5(b)], and this is due to a decrease in IMX density and therefore a reduction of repulsive IMX--IMX interaction. 

For all delay times, $M_{\rm 1}$(x) is peaked in the laser excitation region around $x = 0$ and reduces with 
$x$, following the reducing IMX density away from the laser excitation spot. Figure~5(c) presents the IMX energy shift $M_{\rm 1}(t) - M_{\rm 1}(t=12$~ns) at $x = 0$ (here the IMX energy is plotted relative to the IMX energy at $t = 12$~ns, to compensate for the IMX diamagnetic energy shift~\cite{Kuznetsova17}). A larger blue shift is observed over time for higher magnetic fields [Fig.~5(c)]. This is due to the IMX density accumulation in the excitation spot region originating from the suppression of IMX transport away from the excitation spot with increasing $B$ [Fig.~2(a)]. The suppression of IMX transport and, as a result, IMX density accumulation near the IMX generation site, is stronger with increasing magnetic fields. This effect is also presented in Fig.~5(d) showing that the IMX transport distance $R$ anti-correlates with the IMX energy shift at $x = 0$. 

\begin{figure}[htbp]
\centering
\includegraphics[width=\linewidth]{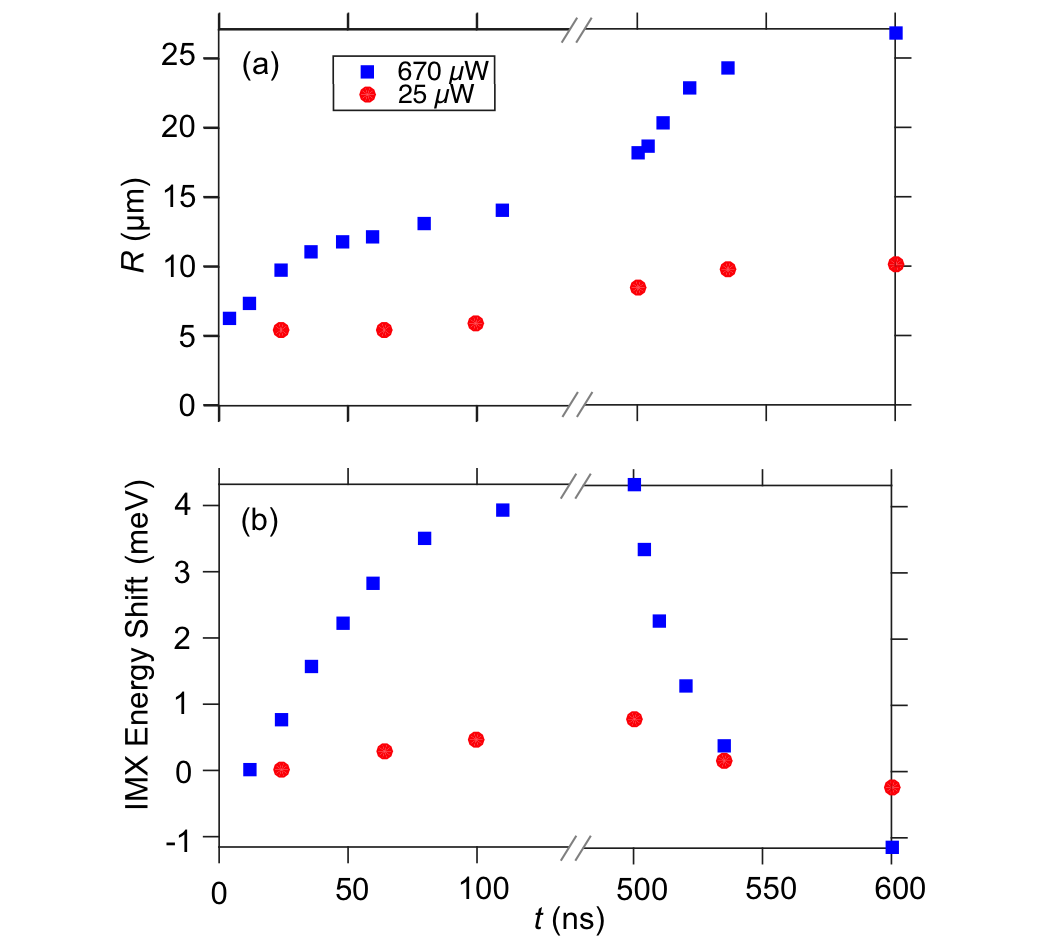}
\caption{Laser excitation power dependence. (a) IMX transport radius $R$ vs. $t$ for average laser excitation powers $P_{\rm ex} = 25$ and 670~$\mu$W. (b) IMX energy shift at $x = 0$ vs. $t$ for $P_{\rm ex} = 25$ and 670~$\mu$W. $T_{\rm bath} = 1.5$~K and $B = 3$~T for all data.}
\end{figure}

Figure~6(a) shows the laser excitation power, $P_{\rm ex}$, dependence of the evolution of the IMX transport distance. The IMX density is controlled with $P_{\rm ex}$; higher $P_{\rm ex}$ corresponds to higher IMX density. At low densities ($P_{\rm ex} =25$~$\mu$W), the radius of IMX emission cloud is essentially that of the laser excitation spot, indicating that IMX transport is suppressed. In contrast, at high densities ($P_{\rm ex} =670$~$\mu$W), the IMX cloud extends well beyond the laser excitation spot due to long-range IMX transport [Fig.~6(a)]. The different IMX transport distances between the two densities is understood in terms of disorder screening. At low densities, IMXs are localized in the disorder potential of the sample. However, since IMXs interact repulsively, at higher densities, they effectively screen the disorder potential and become delocalized, and long-range IMX transport is observed at higher excitation powers.

Figure~6(b) shows that the IMX energy shift is stronger for the higher laser power. A higher repulsive IMX interaction at higher IMX densities contributes to this. We note however that population of higher LL states
can also contribute to a larger blue shift. An investigation of the kinetics of individual LL states is the subject of future work.

\section{III. Theory}

A model based on exciton transport and thermalization is used to simulate the IMX kinetics in high magnetic fields. We solved the following set of coupled equations for the IMX density $n$ and temperature $T$,
\begin{eqnarray}
\frac{\partial n}{\partial t} = \nabla \left[D\nabla n + \mu_{\rm x} n \nabla (u_0 n)\right] + \Lambda - \frac{n}{\tau}\,, \label{transportEqn} \\
\frac{\partial T}{\partial t} = S_{\rm pump} - S_{\rm phonon}\,. \label{thermalizationEqn}
\end{eqnarray}
Here, $\nabla$ is the 2D nabla operator using cylindrical symmetry. The first and second terms in square brackets in equation (\ref{transportEqn}) describe IMX diffusion and drift currents, respectively. The latter originates from the repulsive dipolar interactions and is approximated by $u_0 = 4 \pi e^2 d/\varepsilon$ \cite{Hammack09}. The diffusion coefficient $D(n,T,M)$ is inversely proportional to the IMX effective mass $M(B)$ and describes the magnetic field induced reduction in IMX transport. $D$ also accounts for the screening of the random QW disorder potential by IMXs \cite{Ivanov02, Ivanov06, Hammack09}. The mobility $\mu_{\rm x}$ is given by $\mu_{\rm x} = D(e^{T_0/T} - 1)/(k_BT_0)$ where $T_0 = \pi\hbar^2n/(2M(B)k_B)$ is the quantum degeneracy temperature. The IMX generation rate $\Lambda$ has a Gaussian profile with the same width as the laser excitation beam. $\tau$ is the IMX optical lifetime. The thermalization equation (\ref{thermalizationEqn}) describes heating of excitons by non-resonant photoexcitation, $S_{\rm pump}(T_0,T)$ and cooling via interaction with bulk longitudinal acoustic phonons, $S_{\rm phonon}(T_0,T)$. The emission intensity is extracted from $n/\tau$. Expressions for $D$, $S_{\rm pump}$, $S_{\rm phonon}$ and $\tau$ can be found in \cite{Hammack09}.

\begin{figure}[tbp]
\centering
\includegraphics[width=\linewidth]{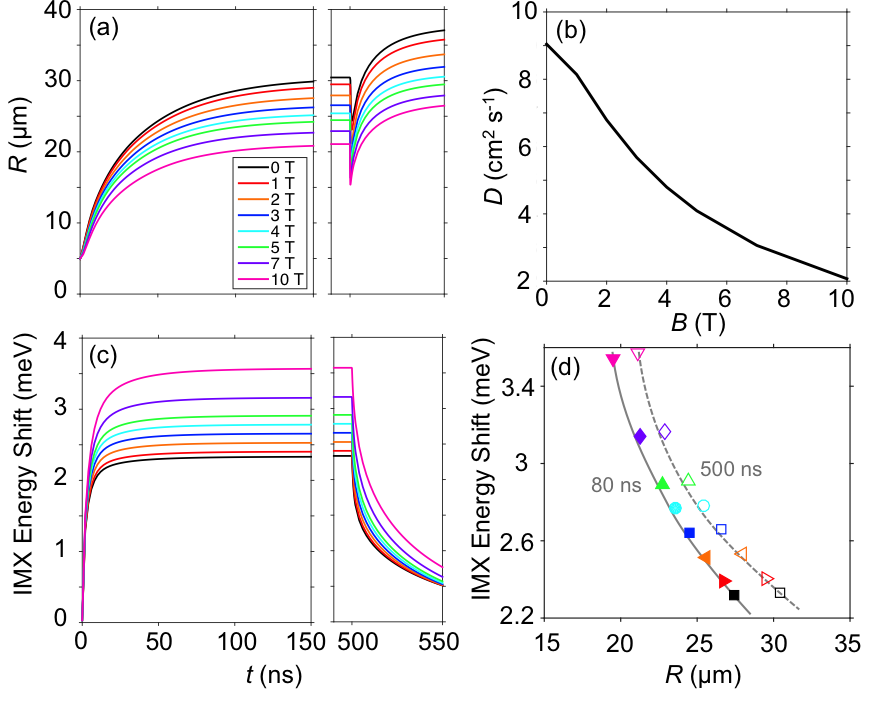}
\caption{(a) Calculated IMX transport distance $R$ vs. delay time $t$. (b) Spatially averaged diffusion coefficient $D$ vs. $B$ for $t = 10$~ns. (c) Calculated IMX energy shift at $x=0$ vs. $t$. (d) IMX energy shift at $x=0$ vs. $R$ for $t = 80$~ns (filled markers) and $t = 500$~ns (open markers).}
\end{figure}

The magnetic field dependence enters equations (\ref{transportEqn}-\ref{thermalizationEqn}) via $D$ and $T_0$ since both depend on $M(B)$. In addition, $\tau$ is given by the single $k=0$ IMX lifetime $\tau_r(B)$ divided by the fraction of IMXs that are inside the radiative zone. The radiative zone is the region in momentum space contained within the intersection between the photon and exciton dispersion surfaces, the latter being a function of $M(B)$~[Fig.~1(b)]]. The increase in $M(B)$ decreases the energy width of the radiative zone which lowers its occupation and enhances $\tau$. At the same time, the magnetic field shrinks the in-plane Bohr radius, increasing the probability of electron-hole recombination and thus decreasing $\tau$. $M(B)$ and $\tau_r(B)$ were determined in Refs.~\cite{Wilkes16, Wilkes17} by calculating eigenstates of the Hamiltonian describing the relative motion of a Coulomb bound electron-hole pair using the multi-sub-level approach~\cite{Wilkes16, Wilkes17, SivalertpornPRB2012}. 

The experimental data in Fig.~4(c) shows the IMX lifetime increasing with $B$. This indicates that the reduction of the radiative zone is the dominant effect in determining the IMX lifetime. In contrast, the calculated $\tau$ decreases with increasing $B$ at temperatures below about 2\,K and increases with increasing $B$ at higher temperatures~\cite{Wilkes17}. This discrepancy is likely due to the non-parabolicity of the exciton dispersion which was not included in the estimates~\cite{Wilkes17}.

\begin{figure}[htbp]
\centering
\includegraphics[width=\linewidth]{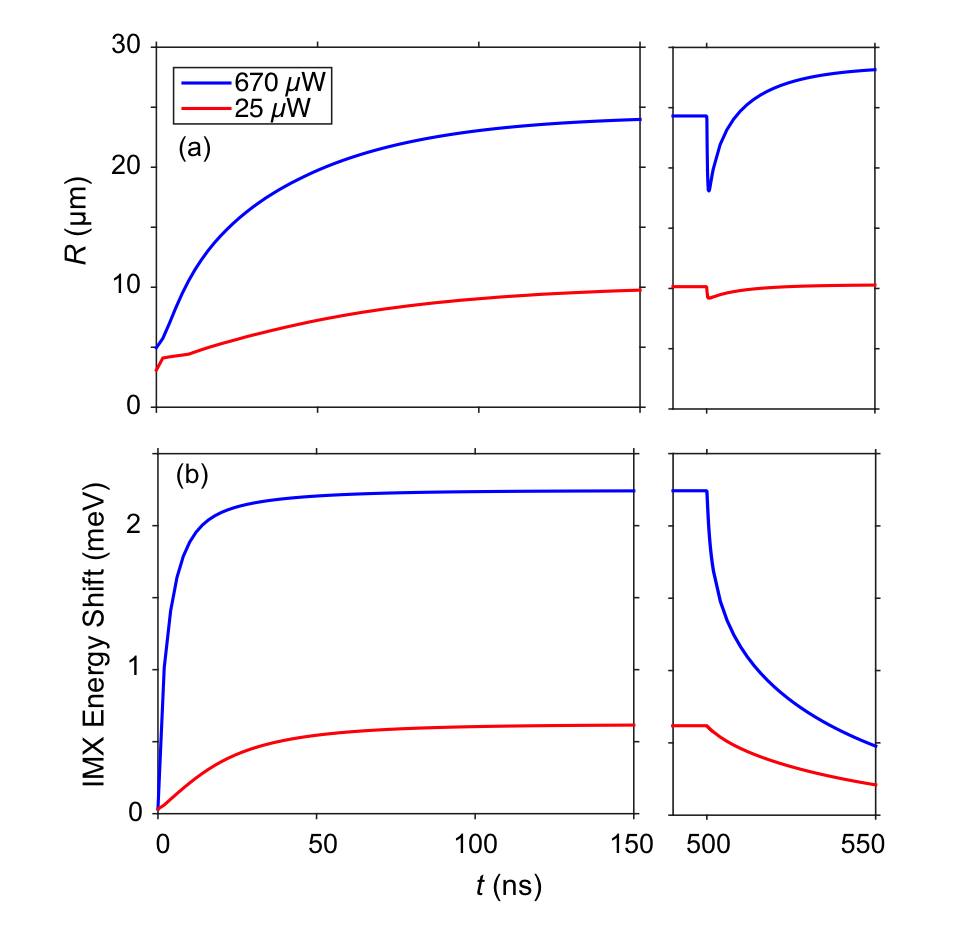}
\caption{Calculated IMX transport distance $R$ (a) and IMX energy shift at $x = 0$ (b) vs. delay time $t$ for $P_{\rm ex} = 25$~$\mu$W (red) and 670 $\mu$W (blue).}
\end{figure}

Figure~7(a) and (c) shows the calculated IMX transport distance and energy shift vs. delay time for all magnetic fields. The theoretical simulations are in qualitative agreement with the experimental data. The spatially averaged diffusion coefficient $D$ plotted in Fig.~7(b) exhibits a similar dependence as the experimentally extracted $D^{*}$ in Fig.~3(c). The evolution of the IMX density $n(r,t)$ from Eq.~1 shows that  
$D^{*} \approx D + \mu_{\rm x}nu_{\rm 0}$. The theoretical model also shows a higher blue shift in IMX PL energy at the origin due to the suppression of IMX transport at higher magnetic fields [compare Fig.~5(d) to Fig.~7(d)]. 

Figure~8(a) and (b) shows the calculated IMX transport distance and energy shift vs. delay time for a high and low laser excitation power at $B = 3$~T. The calculations show longer transport distances and higher energy shift at higher laser power, consistent with the experiment [compare Fig.~6 to Fig.~8]. 

We note that IX PL intensity is suppressed in the excitation spot region during the excitation pulse due to the laser-induced heating of IXs resulting in a lowering of the radiative zone occupation~\cite{Butov02, Ivanov06, Hammack09, Kuznetsova12, Kuznetsova17}. After the excitation pulse termination, IXs cool down, and the IX PL intensity increases at the excitation spot region a few ns after the excitation pulse ends~\cite{Hammack09}. This results in the apparent reduction kink in $R$ [Fig.~7(a)] (the kink is not observed in the experiment within the experimental resolution). An enhancement of $R$ after the excitation pulse termination is also observed in the simulations [Fig.~7(a), 8(a)], consistent with the experiment [compare Fig.~6(a) to Fig.~8(a)]. This enhancement of the IMX transport distance is consistent with a better screening of the QW disorder potential by colder IMXs. Inclusion of population kinetics of higher LL IMX states and their effect on $R$ is the subject of future work.

\section{IV. Summary}

Transport of IMXs in high magnetic fields was studied with time-resolved optical imaging. Slower IMX transport was observed at higher magnetic fields. This was attributed to the exciton effective mass increase with magnetic field. An enhancement of the IMX energy at the laser excitation spot was found to anti-correlate with the IMX transport distance. This was attributed to an accumulation of repulsively interacting IMXs near their generation site due to the suppression of IMX transport at high magnetic fields. Faster IMX transport was observed at a higher laser excitation power. This was attributed to IMX delocalization from the disorder potential due to disorder screening at higher IMX densities. A theoretical model of IMX transport is in agreement with the experiment. 

\section{ACKNOWLEDGMENTS}

These studies of indirect excitons were supported by NSF Grant No. 1640173 and NERC, a subsidiary of SRC, through SRC-NRI Center for Excitonic Devices and by NSF Grant No. 1407277. C.J.D. was supported by the NSF Graduate Research Fellowship Program under Grant No. DGE-1144086. J.W. was supported by the EPSRC (Grant EP/L022990/1).

\end{document}